# 3GPP Evolution from 5G to 6G: A 10-Year Retrospective

Xingqin Lin

NVIDIA

Email: xingqinl@nvidia.com

*Abstract*— The 3rd Generation Partnership Project (3GPP) evolution of mobile communication technologies from 5G to 6G has been a transformative journey spanning a decade, shaped by six releases from Release 15 to Release 20. This article provides a retrospective of this evolution, highlighting the technical advancements, challenges, and milestones that have defined the transition from the foundational 5G era to the emergence of 6G. Starting with Release 15, which marked the birth of 5G and its New Radio (NR) air interface, the journey progressed through Release 16, where 5G was qualified as an International Mobile Telecommunications-2020 (IMT-2020) technology, and Release 17, which expanded 5G into new domains such as non-terrestrial networks. Release 18 ushered in the 5G-Advanced era, incorporating novel technologies like artificial intelligence. Releases 19 and 20 continue this momentum, focusing on commercially driven enhancements while laying the groundwork for the 6G era. This article explores how 3GPP technology evolution has shaped the telecommunications landscape over the past decade, bridging two mobile generations. It concludes with insights into learned lessons, future challenges, and opportunities, offering guidelines on 6G evolution for 2030 and beyond.

## I. INTRODUCTION

The journey of mobile communication technologies has always been a story of relentless innovation, ambition, and collaboration. Over the past decades, the 3rd Generation Partnership Project (3GPP) has been at the forefront of shaping the telecommunications landscape, transforming the possibilities of wireless communication [1]. From the advent of the fifth generation (5G) to the initial steps toward the sixth generation (6G), 3GPP's development timeline as illustrated in Figure 1, spanning Releases 15 to 20, offers a rich roadmap of the latest mobile technology breakthroughs in the last decade. This 10-year retrospective examines how 3GPP has evolved the foundation of mobile networks, enabling a seamless transition from the 5G era to the dawn of 6G.

3GPP started its 5G work by hosting a workshop in September 2015. At the workshop, there was an emerging consensus that there would be a new, non-backward compatible radio access technology (RAT) for 5G. After 5G studies in the Release-14 timeframe, 3GPP completed Release 15 in June 2018, marking the birth of 5G [2]. This milestone introduced the New Radio (NR) air interface, laying the groundwork for a new generation of connectivity defined by improved speed, reliability, and flexibility [3]. Designed to address three International Mobile Telecommunications-2020 (IMT-2020) key usage scenarios (enhanced mobile broadband (eMBB), ultra-reliable low-latency communications (URLLC), and massive machine-type communications (mMTC)) [4], Release 15 established a flexible framework for 5G's diverse applications. It introduced both standalone (SA) and non-standalone (NSA) architectures, enabling a gradual migration from existing Long-Term Evolution (LTE) networks [5]. This release laid the foundation for global 5G deployments, ushering in a new era of enhanced connectivity and enabling a variety of use cases.

Building on the success of Release 15, 3GPP Release 16, finalized in mid-2020, took 5G a step further by qualifying it as an IMT-2020 technology approved by the International Telecommunication Union (ITU) [6]. Release 16 introduced significant enhancements, including support for the industrial Internet of Things (IoT), vehicle-to-everything (V2X) communication, integrated access and backhaul (IAB), operation in unlicensed spectrum, and positioning, among others [7][8]. By broadening 5G's application scope, Release 16 enhanced the network's ability to support emerging use cases, such as smart manufacturing and autonomous vehicles.

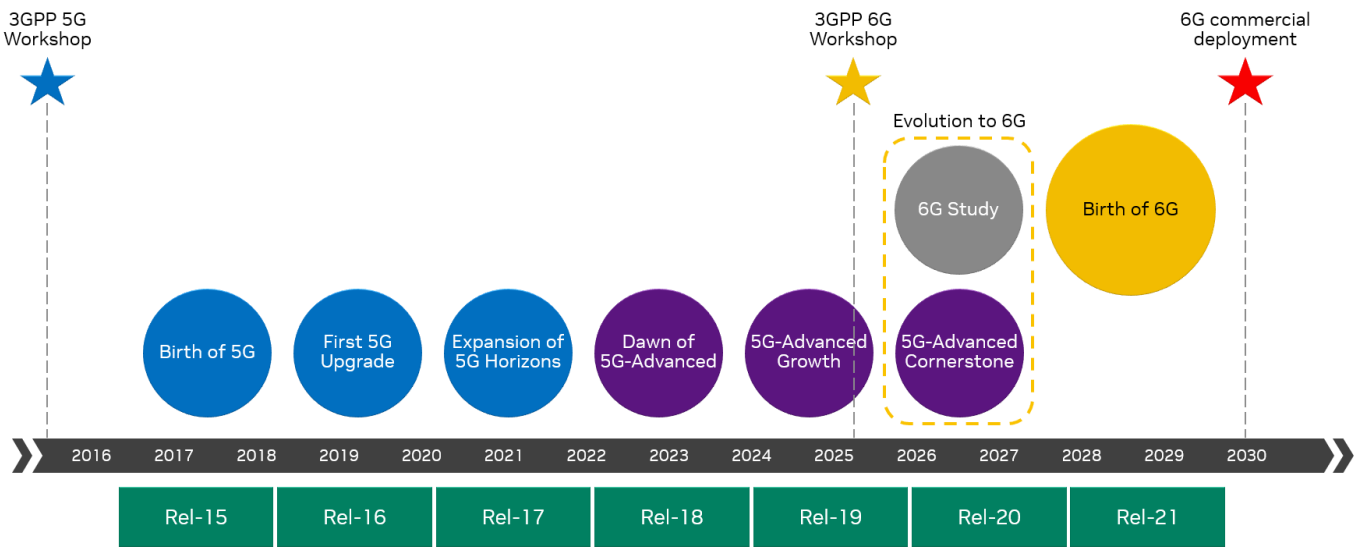

**Figure 1: 3GPP evolution roadmap from 5G to 6G (indicative).**

Release 17, finalized in 2022, represented the expansion of 5G into new domains, pushing the boundaries of what the technology could achieve [9]. It introduced support for non-terrestrial networks (NTNs), allowing connectivity through satellites to extend the reach of 5G to remote and underserved areas. This release also defined reduced capability (RedCap) devices, enabling cost-effective solutions for IoT applications. By continuing to enhance multiple-input multiple-output (MIMO) capabilities, coverage, and power-saving features, Release 17 continued to improve 5G's performance to meet a diverse range of global needs.

A significant inflection point arrived with Release 18, finalized in 2024, marking the beginning of the 5G-Advanced era [10][11]. This phase brought a sharper focus on integrating artificial intelligence (AI)/machine learning (ML) into 5G to optimize use cases such as load balancing and mobility at the 5G architecture level. Release 18 also introduced network energy-saving techniques and enhancements for extended reality (XR) applications, reflecting the growing emphasis on sustainability and immersive experiences. Additionally, Release 18 addressed new deployment needs, such as deployments in a spectrum with less than 5 MHz bandwidth for the future railway mobile communication system (FRMCS).

The momentum of 5G-Advanced continues with Release 19, set for completion in late 2025 [12]. Release 19 integrates AI/ML into the NR air interface, enabling smarter beam management and positioning. It also introduces subband non-overlapping full duplex (SBFD) for more efficient spectrum use, low-power wake-up signal (LP-WUS) to extend device battery life, and ambient IoT, which facilitates ultra-low-power connectivity for massive IoT applications. Additionally, 3GPP conducts studies on channel modeling for integrated sensing and communication (ISAC) and the 7–24 GHz spectrum. These advancements not only enhance 5G-Advanced capabilities but also prepare the evolution toward 6G.

Release 20, set to start in the second half of 2025, represents a critical juncture in the evolution from 5G to 6G. It will serve as both the culmination of 5G-Advanced and the early bridge to 6G. On one hand, it will focus on a more selective set of essential enhancements for addressing commercial deployment needs. On the other hand, it will start a series of 6G studies, covering use cases, service requirements, scenarios, performance requirements, architecture, and technology components. These 6G studies will guide the development of 6G specifications in Release 21, aligning with the ITU's IMT-2030 timeline [13].

Throughout this 10-year journey, 3GPP's evolution has been marked by collaboration among stakeholders. The collective effort has driven advancements not only in 5G performance but also in addressing societal needs such as digital inclusion, sustainability, and public safety [14]. As we approach the 6G era, the lessons learned from 5G's development will serve as valuable guidelines. This article aims to provide a holistic look at the milestones, breakthroughs, and challenges that defined 3GPP's progression from Release 15 to Release 20, as summarized in Table 1. After examining the 3GPP evolution path from 5G to 6G, this article presents five key lessons, offering guidelines on 6G evolution for 2030 and beyond.

## II. 3GPP RELEASE 15: BIRTH OF 5G

3GPP Release 15 marked the inception of the 5G NR system. As the first release dedicated to 5G, it laid the foundational framework for a new era in mobile communication, establishing the technical specifications for both SA and NSA modes of 5G operation. This section provides an overview of 5G design in Release 15. Interested readers may find more detailed information in [15]-[18].

The next-generation radio access network (NG-RAN) for 5G NR systems comprises base stations known as 5G node B (gNB) for NR and next-generation evolved node B (ng-eNB) for LTE connectivity. These components interact via the NG interface for core network communications and the Xn interface for inter-RAN operations. This architecture supports dual connectivity (DC) and network slicing, which allow operators to allocate network resources dynamically, ensuring efficient performance across various applications. Functional splits within the architecture facilitate optimized resource management, seamless session handling, and robust security. The NG-C and NG-U interfaces are designed to enable control plane and user plane communications between RAN and the core network, respectively, while the Xn interface ensures smooth inter-node operations.

Figure 2 provides an illustration of the 5G NR protocol architecture. The physical layer plays a critical role in interfacing the user equipment (UE) with the network. At its core, the NR physical layer adopts orthogonal frequency division multiplexing (OFDM) with a cyclic prefix (CP) as its waveform. This scheme is used for both uplink and downlink transmissions, with an additional option of discrete Fourier transform-spread OFDM (DFT-s-OFDM) for the uplink to improve power efficiency in certain applications. The physical layer is designed to operate across a wide range of spectrum allocations in frequency range 1 (FR1) of 410 MHz – 7.125 GHz and frequency range 2 (FR2) of 24.25 GHz – 52.6 GHz, making it adaptable for global deployments. Resource blocks, each spanning 12 subcarriers, form the building blocks for frequency allocation, and these are managed within a frame structure accompanied by flexible subcarrier spacing of $2^\mu \times 15$ kHz ($\mu = 0, 1, 2, 3$) to support diverse use cases.

In NR, there are several physical channels for both uplink and downlink communication. Downlink channels include the physical downlink shared channel (PDSCH), the physical downlink control channel (PDCCH), and the physical broadcast channel (PBCH). Uplink channels consist of the physical uplink shared channel (PUSCH), the physical uplink control channel (PUCCH), and the physical random access channel (PRACH). These channels are accompanied by reference signals and synchronization signals, ensuring robust connectivity and control within the network.

| 3GPP Release | Enhancements to Existing Features | New Features |
|---|---|---|
| Release 15: Birth of 5G | - N.A. | - 5G New Radio |
| Release 16: First 5G Upgrade | - MIMO<br>- Mobility enhancements<br>- UE power saving<br>- Dual connectivity and carrier aggregation<br>- Cross link interference and remote interference management | - Industrial Internet of Things and ultra-reliable low-latency communications<br>- Unlicensed operation<br>- Sidelink communication<br>- Positioning<br>- Integrated access and backhaul<br>- 2-step random access |
| Release 17: Expansion of 5G Horizons | - MIMO<br>- Coverage enhancements<br>- Industrial Internet of Things and ultra-reliable low-latency communications<br>- Sidelink communication<br>- Positioning<br>- Integrated access and backhaul<br>- Others: UE power saving, dynamic spectrum sharing, multi-subscriber identity module support, small data transmission | - Non-terrestrial networks<br>- Multicast and broadcast services<br>- Reduced capability device support<br>- Operation extension to 71 GHz |
| Release 18: Dawn of 5G-Advanced | - MIMO<br>- Coverage enhancements<br>- Mobility enhancements<br>- Sidelink communication<br>- Positioning<br>- Integrated access and backhaul<br>- Non-terrestrial networks<br>- Multicast and broadcast services<br>- Reduced capability device support<br>- Others: Multi-carrier enhancements, dynamic spectrum sharing, multi-subscriber identity module support, small data transmission | - Network energy savings<br>- Extended reality support<br>- Artificial intelligence/machine learning for next-generation radio access network<br>- In-device coexistence<br>- Uncrewed aerial vehicle<br>- Air-to-ground network<br>- Support for less than 5 MHz spectrum<br>- Network-controlled repeater |
| Release 19: 5G-Advanced Growth | - Multiple-input multiple output (MIMO)<br>- Mobility enhancements<br>- Non-terrestrial networks<br>- Extended reality support<br>- Artificial intelligence/machine learning for next-generation radio access network<br>- Others: sidelink communication, network energy savings, multi-carrier enhancements | - Artificial intelligence/machine learning for NR air interface<br>- Subband non-overlapping full duplex<br>- Low-power wake-up signal/receiver<br>- Ambient IoT<br>- Wireless access backhaul<br>- Femtocell<br>- *Toward 6G*: Channel modeling for integrated sensing and communication and 7-24 GHz spectrum |
| Release 20 Evolution to 6G (Outlook) | - MIMO<br>- Artificial intelligence/machine learning for next-generation radio access network<br>- Artificial intelligence/machine learning for NR air interface<br>- Ambient IoT<br>- Non-terrestrial networks<br>- Integrated sensing and communication | - Artificial intelligence/machine learning for mobility<br>- *Toward 6G*: 6G studies |

Table 1: A summary of 3GPP evolution from Release 15 to Release 20 (non-exhaustive).

For the downlink, the system supports a multitude of quadrature amplitude modulation (QAM) schemes, including quadrature phase shift keying (QPSK), 16-QAM, 64-QAM, and 256-QAM. For the uplink, in addition to the schemes available for the downlink, the system supports $\pi/2$-binary phase shift keying ($\pi/2$-BPSK) for DFT-s-OFDM, providing enhanced power efficiency. To ensure robust and reliable communication, NR employs advanced channel coding techniques. Transport blocks are encoded using low-density parity-check (LDPC) codes, known for their high efficiency and error-correction capabilities. Polar codes are utilized for control channels and the PBCH, leveraging their suitability for short block lengths and latency-sensitive applications.

Key physical layer procedures include cell search, power control, uplink synchronization, and hybrid automatic repeat request (HARQ). These procedures ensure seamless operation in diverse network scenarios, from initial access to maintaining connectivity during mobility. Additionally, channel state information (CSI) reporting and beam management are critical for advanced antenna systems (e.g., massive MIMO) and millimeter wave (mmWave) communications in 5G. Physical layer measurements are integral for ensuring optimal network performance. These include intra- and inter-frequency measurements for handovers, inter-RAT measurements, and timing synchronization. Such measurements provide the basis for advanced radio resource management (RRM) and seamless mobility.

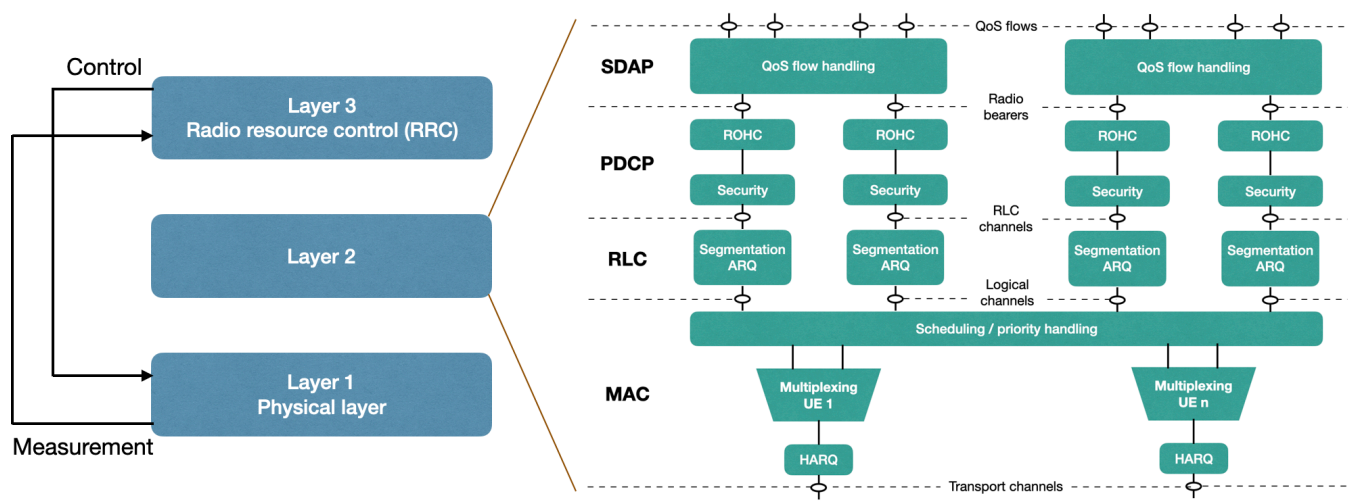

**Figure 2: An illustration of 5G NR protocol architecture.**

On top of the physical layer, the NG-RAN leverages a multi-layered protocol architecture to manage data transmission efficiently. The medium access control (MAC) layer handles logical-to-physical channel mapping and prioritization of traffic, ensuring smooth data flow. The radio link control (RLC) layer offers different transmission modes: transparent, unacknowledged, and acknowledged, catering to diverse reliability requirements. The packet data convergence protocol (PDCP) layer adds security features, robust header compression (ROHC), and sequence to optimize data delivery. The service data adaption protocol (SDAP) layer aligns quality-of-service (QoS) flows with radio bearers, providing an essential link between application requirements and network resource management.

The radio resource control (RRC) layer governs the establishment and maintenance of RRC connections, configuring data and signaling bearers while managing mobility and QoS. It defines RRC_IDLE, RRC_INACTIVE, and RRC_CONNECTED states, allowing devices to operate with varying levels of activity and energy efficiency. These states ensure that devices remain reachable while optimizing power consumption and resource usage. RRC also supports essential mobility features such as handovers and cell reselection, ensuring seamless transitions across cells.

Mobility management is a cornerstone of NG-RAN functionality, enabling devices to maintain continuous connectivity while on the move. Intra-NR mobility supports handovers and beam failure recovery, while inter-RAT mobility facilitates transitions between NR and legacy LTE networks. Paging mechanisms and RAN-based notification area (RNA) updates enhance device reachability. These features collectively ensure that mobility is seamless and user experiences remain uninterrupted.

The QoS framework in NG-RAN is designed to deliver consistent performance across diverse applications. It incorporates mechanisms for adaptive rate control and explicit congestion notifications, which optimize network performance under varying traffic conditions. Additionally, NG-RAN implements a robust security framework to protect user data and signaling communications. The security mechanisms include device and user authentication, encryption, and integrity protection, enabling secure operations across multiple network domains.

NG-RAN's advanced features further improve its performance. Carrier aggregation (CA) allows the combination of multiple carriers to achieve higher data throughput. Beamforming and MIMO technologies improve signal quality and spectral efficiency, enabling better coverage and capacity. Network slicing provides logical partitioning of the network, allowing operators to tailor services for specific use cases such as IoT, critical communications, or high-speed broadband.

In summary, Release 15 of 3GPP marked the "birth of 5G," setting the stage for its evolution. It introduced a flexible design capable of addressing the diverse and demanding requirements of modern mobile communications. This release established the foundation for the subsequent enhancements and expansions that drive the 5G and beyond ecosystem.

### III. 3GPP RELEASE 16: FIRST 5G UPGRADE

Known as the "first 5G upgrade," 3GPP Release 16 significantly expanded the scope and functionality of 5G by enhancing existing features, addressing new use cases, and enabling greater flexibility and efficiency for various industries.

*A. Enhancements to Existing Features*

MIMO, a cornerstone of 5G's ability to deliver high data rates and spectral efficiency, received substantial upgrades in Release 16 [19][20]. Release-15 MIMO faced challenges such as high uplink overhead due to the Type-II CSI reporting mechanism. Release 16 addressed these limitations by introducing compression techniques in both spatial and frequency domains for CSI reporting. These enhancements reduced reporting payloads significantly while preserving spectral efficiency. The introduction of default spatial relation configurations simplified beam management [21]. Additionally, support for multi-transmission-reception-point (multi-TRP) scenarios enhanced both downlink reliability and data rates. Additionally, Release 16 introduced new designs for demodulation reference signal (DMRS) to reduce peak-to-average power ratio.

Mobility management also saw notable improvements in Release 16. Conditional handover (CHO) and dual active protocol stack handover were introduced to address the limitations of traditional handover mechanisms [22]. These features reduced connection interruptions and improved handover reliability, ensuring seamless transitions for UEs between cells. This was particularly important for high-mobility scenarios, such as vehicles and trains, where maintaining uninterrupted connectivity is a significant challenge.

Energy saving was another area where Release 16 delivered meaningful advancements, particularly for UEs [23][24]. In both connected and idle states, UEs benefited from enhanced techniques such as discontinuous reception (DRX) adaptation, cross-slot scheduling, and reduced RRM measurement requirements. These improvements allowed UEs to communicate their power-saving preferences dynamically, optimizing energy consumption while maintaining network performance.

DC/CA, which are essential for maximizing spectrum utilization and ensuring robust network performance, were further improved in Release 16 [25]. Enhancements included asynchronous NR-DC support, direct secondary cell group (SCG)/secondary cell (SCell) configurations, and the introduction of NR SCell dormancy. These changes reduced setup times, enabling better resource allocation and enhancing the overall network experience.

Another area of improvement was interference management, an issue that becomes particularly challenging in dynamic environments and dense deployments. Cross-link interference (CLI) mitigation emerged as an important issue for time-division duplexing (TDD) systems, reducing interference between uplink and downlink transmissions [26]. Remote interference management (RIM) tackled issues caused by atmospheric ducts in wide-area TDD deployments, enabling automated resolution of interference scenarios [27].

*B. New Features*

The industrial IoT emerges as a key vertical supported by Release 16 [28]. Expanding on Release 15's capabilities, Release 16 introduced mechanisms to achieve URLLC for factory automation, energy distribution, and remote industrial applications [29]. Features like uplink preemption allowed higher-priority transmissions to override lower-priority ones in real-time, optimizing latency-sensitive operations. Enhanced collision resolution techniques improved resource efficiency by intelligently handling conflicts between multiple uplink transmissions, ensuring seamless integration of diverse traffic flows. Time-sensitive networking (TSN) was a notable addition for industrial use cases, providing precise synchronization and prioritization of traffic flows. These enhancements collectively enabled NR to meet the demands of industrial automation,

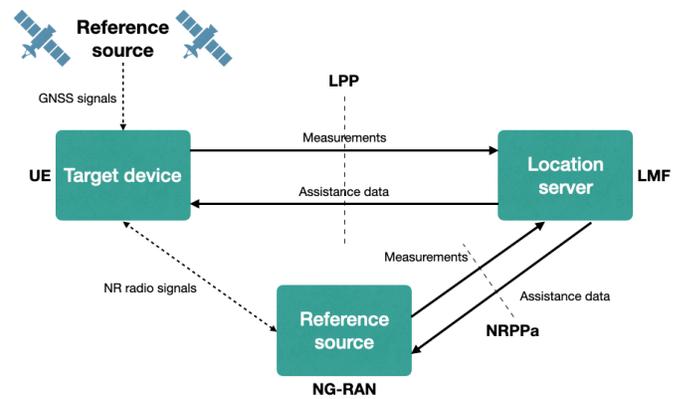

**Figure 3: An illustration of 5G NR positioning framework.**

ensuring reliability and efficiency for mission-critical applications.

Another significant advancement was NR's capability to operate in the unlicensed spectrum, building on the legacy of LTE's licensed assisted access (LAA) [30]. Unlike LTE-LAA, Release 16 enabled standalone operation in unlicensed bands, including the newly allocated 6 GHz spectrum. This functionality enhanced deployment flexibility and increased data rates, using mechanisms such as listen-before-talk (LBT) to ensure harmonious coexistence with other wireless technologies [31].

Intelligent transportation system (ITS) and V2X communication also benefited from Release 16's innovations [32][33]. With sidelink communication as a core component, Release 16 NR supported unicast, groupcast, and broadcast transmissions tailored to ITS applications [34][35]. These enhancements addressed advanced use cases such as vehicle platooning, cooperative lane merging, and remote driving. The integration of robust QoS frameworks ensured that these applications operate reliably under diverse conditions.

Accurate positioning capabilities are increasingly essential, especially for indoor applications and emergency services [36]. Figure 3 provides an illustration of the 5G NR positioning framework, where LTE positioning protocol (LPP) is used between UE and location management function (LMF) and NR positioning protocol A (NRPPa) is used between NG-RAN node and LMF [37]. Release 16 introduced advanced positioning methods that complemented traditional global navigation satellite systems (GNSS) [38]. The new positioning reference signal (PRS) allowed for improved time-of-arrival estimation, enabling precise positioning even in challenging environments. Uplink-based positioning enhancements leveraged extended sounding reference signal (SRS) and beamforming to increase accuracy and range.

Release 16 introduced IAB, which allows nodes to serve as both access points and backhaul units [39]. This feature enabled 5G NR to use wireless backhaul as an alternative to traditional fiber links, simplifying the deployment of small cells in dense urban areas and temporary event settings. Operating efficiently in TDD mode and leveraging mmWave bands, IAB facilitates both in-band and out-of-band operations to optimize network architecture and performance [40][41].

Lastly, following a comprehensive study on non-orthogonal multiple access (NOMA) [42], 3GPP conducted minimal normative work on 2-step random access channel (RACH). The 2-step RACH procedure consists of two steps: in message A, the UE transmits a preamble combined with its initial uplink data in a single transmission; in message B: the gNB responds with a combined random-access response and contention resolution message, which includes information about resource allocation and acknowledgment of the UE's request [43]. Though 2-step RACH has the potential of latency reduction, it is faced with implementation complexity.

## IV. 3GPP RELEASE 17: EXPANSION OF 5G HORIZONS

3GPP Release 17 further expanded 5G horizons by advancing existing features while introducing innovations tailored to emerging verticals and use cases.

### A. Enhancements to Existing Features

MIMO was further enhanced in Release 17 to improve spectral efficiency, latency, and reliability. Key enhancements included a unified transmission configuration indicator (TCI) framework for uplink and downlink beam management. Multi-TRP support was extended to PDCCH, PUCCH, and PUSCH. Enhanced SRS configurations enabled better uplink and downlink CSI acquisition, supporting more antennas and improving coverage. Additionally, CSI reporting enhancements leveraged angle-delay reciprocity for improving accuracy and reducing UE computation. These advancements enhanced MIMO's applicability across diverse deployment scenarios, including high-speed trains and challenging propagation environments.

NR coverage was enhanced in Release 17 [44]. These enhancements included improvements to PUSCH repetition, providing better coverage with dynamic slot-based repetition and advanced PUSCH dropping rules. Transport block processing over multiple slots was another major advancement, enabling a single transport block to be transmitted across multiple slots. This approach lowered coding rates and increased energy per resource element, leading to enhanced coverage. Additionally, DMRS bundling was introduced to improve channel estimation for PUSCH and PUCCH. The dynamic PUCCH repetition factor mechanism allowed the network to adapt repetition factors based on channel conditions, improving efficiency. These features collectively strengthened NR's ability to deliver reliable connectivity in diverse scenarios.

In Release 17, 3GPP enhanced NR's support for industrial IoT and URLLC to further address stringent latency and reliability demands across verticals. Key enhancements included improved HARQ and CSI feedback mechanisms for higher reliability, prioritization of traffic with varying priorities for efficient intra-UE multiplexing, and support for URLLC in shared spectrum via semi-static channel occupancy. Time synchronization for TSN was enhanced with propagation delay compensation methods. Non-public networks were also enhanced, e.g., broadcasting network selection information enabled UEs to connect to a non-public network efficiently without impacting cell selection or mobility [45].

Release 17 introduced NR sidelink enhancements aimed at improving power efficiency, reliability, and latency. Key features included power-saving resource allocation methods such as partial sensing and random resource selection. Inter-UE coordination enabled resource optimization through information exchange between devices. Sidelink DRX supported energy-efficient operation with configurable timers for unicast, groupcast, and broadcast communication. Additionally, Uu and sidelink DRX alignment was supported to improve synchronization. Sidelink relay functionality was enhanced for single-hop UE-to-network relay, allowing remote UEs to connect via relay UEs under various scenarios. These updates strengthened sidelink's role in public safety, IoT, and vehicular applications.

NR positioning was further enhanced to improve accuracy, latency, and reliability. Key advancements included refined timing error mitigation for positioning measurements, and multi-path mitigation. Latency was reduced through preconfigured measurement gaps, PRS processing windows, and faster sampling. On-demand PRS transmission enabled dynamic control of positioning signals, while support for RRC_INACTIVE allowed efficient, low-power positioning. These enhancements facilitated robust, precise, and efficient positioning for advanced 5G applications and beyond.

Release 17 enhanced IAB by improving robustness, load balancing, spectral efficiency, and backhaul performance. Key enhancements included inter-donor migration for better topology management, reduced service interruptions during node migrations, and new mechanisms for uplink re-routing during link failures. Duplexing enhancements enabled simultaneous transmission and reception, optimizing spectrum use. Scheduling and congestion control improvements addressed end-to-end performance and efficiency. These enhancements made IAB more resilient, efficient, and applicable across FR1 and FR2.

Other enhancements to existing features in Release 17 included:

- UE power savings: In idle/inactive mode, new paging mechanisms reduced false alarms by subdividing UE groups and leveraging paging early indications. In connected mode, power-saving measures included reduced PDCCH monitoring with dynamic adaptations, search space set group switching, and PDCCH skipping.
- Dynamic spectrum sharing (DSS): DSS offers an efficient migration path from LTE to NR by enabling both technologies to share the same carrier [46]. In Release 17, cross-carrier scheduling was enhanced for more flexible DSS.
- Multi-subscriber identity module (SIM) support: Release 17 introduced enhancements for multi-SIM devices to address paging collision and network switching issues [47].
- Small data transmission: Release 17 enabled UEs to transmit small data packets while remaining in the RRC_INACTIVE state, reducing the need for frequent transitions to RRC_CONNECTED.

### B. New Features

Release 17 introduced NTN by integrating satellites and high-altitude platforms into 5G [48][49]. Figure 4 provides an illustration of 5G NTN architecture. Key enhancements in Release 17 addressed challenges such as long latency, Doppler shifts, and large cell sizes. NTN supported mobility across terrestrial and non-terrestrial systems, with enhanced synchronization, handover, and signaling mechanisms. Flexible service link configurations enabled Earth-fixed, quasi-Earth-fixed, and Earth-moving cell coverage. Enhanced radio frequency (RF) and RRM ensured seamless connectivity for NTN-capable devices. With support for geosynchronous (GSO) and non-geosynchronous (NGSO) satellites, NTN can bridge gaps where terrestrial networks (TNs) are impractical or unavailable.

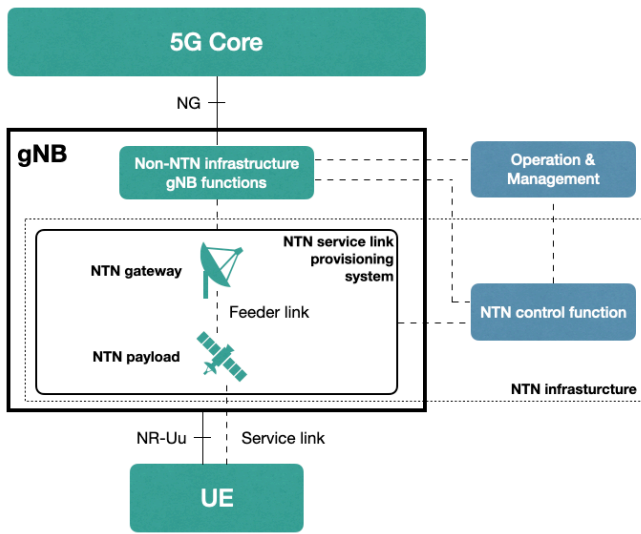

Figure 4: An illustration of 5G NTN architecture.

NR multicast and broadcast services (MBS), introduced in Release 17, offered efficient point-to-multipoint data delivery modes, addressing use cases like public safety, V2X, live video, and IoT applications [50]. The MBS multicast mode supported high-QoS services for UEs in RRC_CONNECTED, utilizing group scheduling and HARQ feedback. Dynamic point-to-point and point-to-multipoint switching optimized scheduling based on QoS and link conditions. Lossless handovers ensured data continuity for multicast sessions. The MBS broadcast mode catered to lower-QoS services, offering downlink-only delivery for UEs in all RRC states [51]. Features like MBS frequency prioritization and interest indications supported service continuity, enhancing system efficiency and user experience.

RedCap device support was introduced in Release 17, enabling low-complexity and power-efficient operations for IoT applications such as industrial sensors, wearables, and surveillance [52]. RedCap UEs supported reduced bandwidths (20 MHz in FR1, 100 MHz in FR2), limited antenna configurations, and fewer downlink MIMO layers. They excluded features like CA and DC. Enhanced DRX cycles and RRM measurement relaxation reduced power consumption, especially in idle/inactive states. Charging differentiation was supported via new requirements for core network functions.

In Release 17, 3GPP extended NR operation to 71 GHz, introducing a new unlicensed band, n263 (57–71 GHz). Accordingly, FR2 frequency range was revised to be 24.25 GHz – 71 GHz, and further divided into FR2-1 and FR2-2, covering 24.25 GHz to 52.6 GHz and 52.6 GHz to 71 GHz, respectively. Subcarrier spacings of 120 kHz, 480 kHz, and 960 kHz were supported, enabling channel bandwidths up to 2 GHz. Features included dynamic spectrum access (LBT and non-LBT modes), beam management, multi-slot PDCCH monitoring, enhanced PUCCH formats, and HARQ timing optimizations. CA with FR1 anchors and intra-band contiguous aggregation were supported.

## V. 3GPP RELEASE 18: DAWN OF 5G-ADVANCED

3GPP Release 18, marking the beginning of 5G-Advanced, represented a significant leap forward in both enhancing existing technologies and addressing emerging verticals and deployment scenarios.

### A. Enhancements to Existing Features

The MIMO evolution in Release 18 focused on enhancing performance to address deficiencies in signaling latency, overhead, spectral efficiency, and coverage. Key improvements included multi-TRP support, enhanced codebooks, and expanded spatial multiplexing capabilities. Multi-TRP support extended the unified TCI framework to allow simultaneous communication with multiple TRPs, optimizing beam indication and timing advance adjustments. The enhancements facilitated efficient resource allocation and dynamic TRP switching. Enhanced codebooks supported an increased number of ports, coherent joint transmission, and dynamic TRP selection. This improved multi-user MIMO performance, especially for high/medium-speed scenarios. For spatial multiplexing, enhanced DMRS configurations doubled the number of orthogonal ports for both uplink and downlink, increasing multi-user MIMO scheduling capacity. Uplink capabilities supported up to 8 transmission layers with advanced codebook and non-codebook options for high-end UEs, e.g., customer premises equipment and industrial devices.

In Release 18, further NR coverage enhancements built on Release-17 efforts to address remaining coverage challenges for bottleneck channels, introducing improvements in PRACH coverage and dynamic waveform switching, among others. PRACH coverage enhancements included support for repetitions in the 4-step RACH procedure, enabling UEs to dynamically adjust repetition numbers based on synchronization signal block (SSB) reference signal received power thresholds. Dynamic waveform switching between DFT-s-OFDM and CP-OFDM was enabled, allowing networks to optimize waveforms dynamically for coverage or capacity without requiring RRC reconfiguration.

NR mobility enhancements in Release 19 focused on reducing latency, enhancing reliability, and optimizing inter-cell and DC transitions. Key advancements included layer 1/layer2 triggered mobility (LTM) with low-latency procedures, selective activation of cell groups in NR-DC, and CHO enhancements. Figure 5 provides an illustration which compares LTM to conventional layer 3 based mobility. Besides, New Release-17 configurations enabled RACH-less mobility, early timing acquisition, and efficient cell reselection using idle/inactive mode measurement results. Enhanced CHO mechanisms supported simultaneous primary cell (PCell) and primary secondary cell (PSCell) evaluation.

Release 18 introduced further advancements in NR sidelink, enhancing its applications for V2X communication. Among the key enhancements was the introduction of CA in the ITS spectrum. Packet duplication ensured robust transmissions across aggregated carriers, benefiting both in-coverage and out-of-coverage UEs. NR sidelink was also extended to operate in unlicensed spectrum bands. Advanced channel access mechanisms were adopted, along with multi-channel access procedures. These advancements facilitated reliable communication in the unlicensed spectrum while maintaining compatibility with existing sidelink operations. Release 18 also addressed LTE-NR sidelink coexistence in the 5.9 GHz spectrum, enabling dynamic resource sharing between LTE and NR sidelink. Relay capabilities were enhanced to extend connectivity and service continuity. Single-hop UE-to-UE and UE-to-network relay support expanded coverage for out-of-coverage scenarios. Multi-path relay functionality enabled UEs to communicate simultaneously over direct and indirect paths.

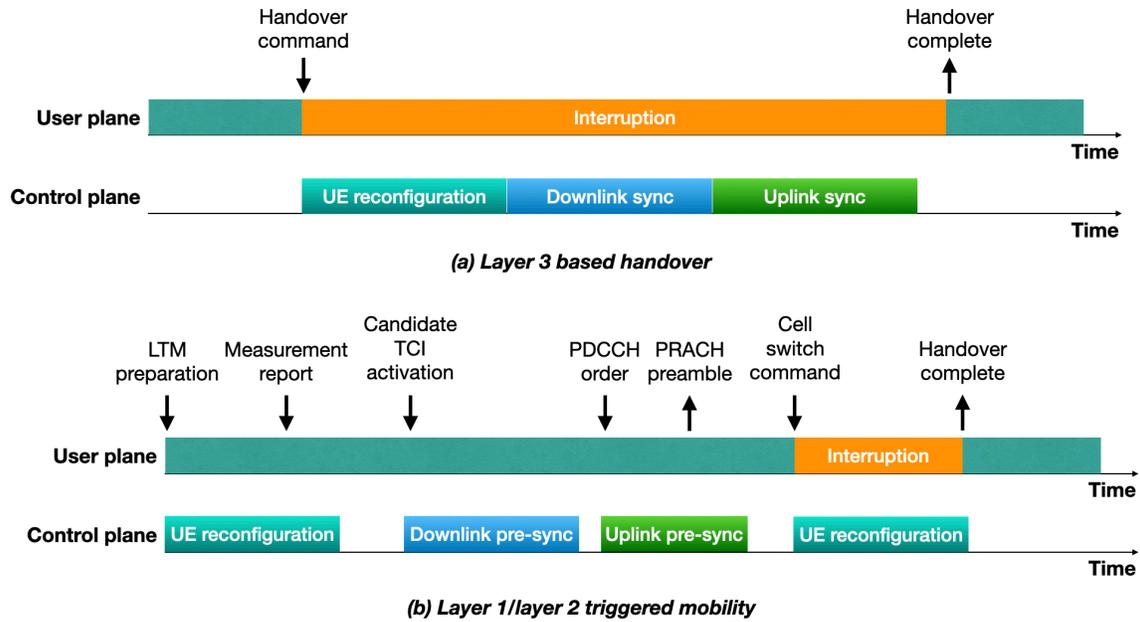

**Figure 5:** An illustration of layer 3 based mobility vs. LTM.

Another major enhancement area in Release 18 was NR positioning, introducing new features for sidelink positioning. The work also introduced carrier phase positioning for centimeter-level precision, utilizing carrier phase measurements for both downlink and uplink methods. Further enhancements addressed RedCap UEs, incorporating frequency hopping for wider bandwidth measurements. To ensure robust positioning, Release 18 included positioning integrity frameworks, alignment with regulatory requirements, and mechanisms for collision avoidance.

The Release-18 work on IAB extended Release-17 developments to support mobile IAB nodes. Mobile IAB nodes, mounted on vehicles, act as RAN nodes, providing NR access links for UEs and backhaul links to parent nodes. Mobile IAB supported integration with different IAB donors, ensuring seamless transitions during migrations. Mechanisms like collision mitigation for physical cell identity and RACH configurations were employed. Enhancements for Rel-18 UEs included RACH-less handovers during IAB-DU migrations and prioritization of mobile IAB cells for inter-frequency reselection.

The NR NTN enhancements in Release 18 improved uplink coverage, supported deployment scenarios in FR2 bands, and enabled mobility and service continuity in NTN environments. Key features included uplink enhancements such as PUCCH repetition for HARQ and NTN-specific DMRS bundling, and support for NTN operation in 17.3–30 GHz using frequency division duplex (FDD). Mobility management was enhanced with NTN-to-TN and TN-to-NTN transitions, RACH-less handovers, and CHO for dynamic environments. Network-verified UE location with round-trip time based positioning supported regulatory needs. Additionally, a 30 MHz channel bandwidth was introduced for FR1 bands, improving NTN operational capabilities for diverse applications.

Release-18 enhancements to MBS built on the Release-17 features to support multicast reception for UEs in RRC_INACTIVE state, enable shared processing for broadcast and unicast reception, and improve resource efficiency in RAN sharing scenarios. Multicast reception for UEs in RRC_INACTIVE state allowed a large number of UEs to access services like mission critical services efficiently. Shared processing enabled UEs to simultaneously receive broadcast and unicast services while optimizing hardware resources. To enhance resource efficiency in RAN sharing scenarios, mechanisms were introduced to avoid duplicated resource allocation for identical MBS content provided by different operators.

RedCap enhancements were introduced in Release 18 for further complexity reduction and power saving. Complexity reduction included peak data rate limit of 10 Mbps and baseband bandwidth reduction to 5 MHz, enabling simpler and more cost-effective devices. Enhanced power savings were achieved through further extended DRX cycles in RRC_INACTIVE state, matching the extended cycles in RRC_IDLE state. These features made RedCap devices suitable for a broader range of applications.

Other enhancements to existing features in Release 18 included:
- Multi-carrier enhancements: This work focused on improving efficiency and performance in multi-cell and multi-band scenarios. One enhancement was multi-cell PDSCH/PUSCH scheduling with a single downlink control information (DCI).
- DSS: Key improvements included enabling UE reception of NR PDCCH candidates overlapping with LTE cell-specific reference signal resource elements, addressing the PDCCH capacity bottleneck caused by increasing NR traffic.
- Multi-SIM support: Release 18 improved NR support for devices managing dual active connections, by addressing temporary hardware conflicts arising when device resources are shared between two active SIMs.
- Small data transmission: Release 18 enhanced small data transmission by enabling downlink-triggered small data delivery for UEs in RRC_INACTIVE state, which avoids transitions to RRC_CONNECTED for small, infrequent packets.

*B. New Features*

The NR network energy savings enhancements introduced in Release 18 aimed to reduce environmental impact and operational costs by optimizing energy consumption in 5G networks [53]. Key features included CSI enhancements for spatial and power domain adaptations, cell discontinuous transmission (DTX)/DRX mechanisms enabling periodic active and inactive periods for gNB, and SSB-less SCell operation for inter-band CA [54]. A barring mechanism was introduced to prevent legacy UEs from camping on cells using energy-saving techniques. Inter-node beam activation facilitated selective SSB beam switching and enhanced paging optimized beam usage for stationary UEs, improving network energy efficiency.

XR enhancements in Release 18 focused on improving awareness, power saving, and capacity to better support XR services [55]. For XR awareness, the enhancements included improved uplink scheduling through additional buffer size tables, delay status reports, and uplink assistance information (e.g., jitter, burst arrival, and periodicity) reporting per QoS flow. For power saving, gNBs can configure DRX cycles aligned with video frame rates (e.g., 30, 60 frames per second) and enable configured grants without monitoring uplink retransmissions [56]. For capacity, enhancements included multiple configured grant PUSCH transmission occasions within a single period and signaling unused configured grant PUSCH occasions via uplink control information. Protocol data unit (PDU) set based discard mechanisms enabled efficient packet handling, discarding redundant PDUs when no longer needed. In congestion scenarios, PDU importance can dictate discard prioritization, with shorter discard timers for lower-priority service data units.

The Release-18 work on AI/ML for NG-RAN enhanced data collection and signaling within NG-RAN to support AI/ML-based functions, including network energy saving, load balancing, and mobility optimization [57][58]. It introduced signaling procedures for data exchange, enabling AI/ML model training and inference in either NG-RAN nodes or operation, administration, and management (OAM) systems. For energy saving, AI/ML optimized cell/node activation to reduce consumption, utilizing an "energy cost" metric. Load balancing leveraged AI/ML predictions and feedback to distribute traffic efficiently across cells. Mobility optimization improved handover success by using UE trajectory predictions and measured trajectory feedback.

In-device coexistence enhancements address internal interference within UEs between a 3GPP RAT and other RATs like WiFi, Bluetooth, and ultra-wideband (UWB) [59]. The Release-18 work introduced frequency division multiplexing (FDM), time division multiplexing (TDM), and autonomous denial solutions to mitigate interference between NR and other RATs within UEs. The FDM solution allowed UEs to report affected NR frequencies. The TDM solution enabled UEs to propose preferred active durations and cycles for affected frequencies, optimizing scheduling to minimize conflicts. The autonomous denial solution allowed UEs to selectively deny uplink transmissions within limits.

Release 18 enhanced NR support for uncrewed aerial vehicles (UAVs), building on previous LTE-based solutions [60][61]. These enhancements addressed UAV-specific requirements, such as low latency for control, high data rates for multimedia, and interference management to protect terrestrial networks. Key features included altitude reporting, altitude-dependent event triggers for interference reporting, and improved flight path reporting with dynamic updates. NR also introduced 'broadcast remote identification' and 'detect and avoid' using sidelink, aerial-specific out-of-band emission compliance, and subscription-based aerial UE identification.

Air-to-ground (ATG) networks enabled in-flight connectivity using ground-based cell towers to communicate with ATG terminals mounted on aircraft [62]. The Releaes-18 work focused on adjacent channel coexistence, RF requirements, and RRM for ATG deployment. Coexistence studies concluded that legacy FR1 TN RF requirements can be reused for ATG systems with defined isolation distances. RF requirements included power control, spurious emissions, and receiver sensitivity for both ATG base stations and UEs. RRM requirements supported enhanced cell reselection, handover delays, timing adjustments, and measurement capabilities tailored for ATG scenarios. These specifications ensured reliable and efficient ATG operations while maintaining coexistence with TNs.

In Release 18, 3GPP extended NR support for dedicated spectrum with less than 5 MHz bandwidth. Specifically, it supported 3 MHz and 5 MHz channels in FR1, facilitating migration from Global System for Mobile Communications–Railway (GSM-R) to 5G NR while accommodating GSM-R carriers [63]. Physical channel adaptations reduced physical resource block (PRB) usage, with PBCH bandwidth limited to 12 PRBs and sizes of control resource set (CORESET) #0 tailored for 12, 15, and 20 PRBs. Short PRACH formats (15 kHz subcarrier spacing) and long PRACH formats (1.25 kHz subcarrier spacing) were supported, with configurations tailored to small bandwidths.

The Release-18 work on network-controlled repeater (NCR) enhanced conventional RF repeaters by allowing them to receive and process side control information from the network [64]. The NCR consisted of two components: one communicating with the gNB via the Uu interface to receive control information, and the other amplifying and forwarding signals between the gNB and UE. Key functionalities introduced in Release 18 included beamforming management, uplink-downlink TDD operation, and ON-OFF state based on control link signals. ON-OFF operation ensures amplification is active only during specified time resources or deactivated upon beam failure. Repeater management supported NCR identification and authorization via gNB and core network coordination.

## VI. 3GPP RELEASE 19: 5G-ADVANCED GROWTH

5G-Advanced evolution continues in ongoing Release 19, enhancing a set of commercially driven features and introducing a suite of innovative features that enhance the capabilities of 5G. Release 19 acts as a bridge toward 6G by conducting forward-looking studies on channel modeling.

### A. Enhancements to Existing Features

MIMO remains a cornerstone of 5G technology, and Release 19 introduces the fifth phase of its evolution. To improve the accuracy and efficiency of beam management, the release enables UE-initiated beam reporting, which allows UEs to trigger reporting rather than relying on gNB requests. Another significant enhancement is the expansion of CSI reporting from 32 to 128 ports, enabling better support for larger antenna arrays, a critical feature for scaling MIMO systems in high-capacity scenarios. Coherent joint transmission capabilities are enhanced to address challenges in scenarios with non-ideal synchronization and backhaul (e.g., inter-site coherent joint transmission). New measurement and reporting mechanisms are introduced for time misalignment and frequency/phase offsets between TRPs. To further improve uplink throughput, the release incorporates enhancements to non-coherent uplink codebooks for UEs equipped with three transmit antennas. Additionally, asymmetric configurations are supported, where a UE receives downlink transmissions from a macro gNB while transmitting to multiple micro TRPs in the uplink. These configurations include enhanced power control mechanisms and path loss adjustments to optimize performance in heterogeneous network environments.

Mobility management is another focus area in Release 19, particularly through the extension of LTM. Initially introduced in Release 18 for intra-central unit (CU) mobility, the new enhancements extend support to inter-CU mobility, enabling smoother transitions across cells associated with different CUs. To further optimize mobility, Release 19 introduces conditional LTM. This combines the benefits of reduced interruption times from LTM with the reliability of CHO. Additionally, event-triggered layer 1 measurement reporting reduces signaling overhead compared to periodic reporting. The inclusion of CSI reference signal (CSI-RS) measurements alongside SSB measurements enhances the performance of mobility procedures.

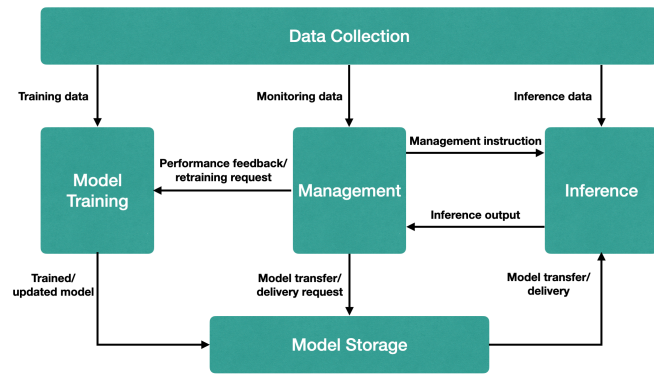

**Figure 6: An illustration of 3GPP AI/ML framework for NR.**

NR NTN evolution continues in Release 19, where 3GPP defines new reference satellite payload parameters to address scenarios with reduced equivalent isotropic radiated power (EIRP) densities per satellite beam compared to previous releases. To accommodate the reduced EIRP, the release explores downlink coverage improvements. Given the large number of UEs expected within a satellite's coverage area, Release 19 also focuses on increasing uplink capacity by incorporating orthogonal cover codes into the DFT-s-OFDM based PUSCH. To support MBS within NTNs, 3GPP enhances MBS by defining signaling mechanisms for specifying intended service areas. Another significant advancement in Release 19 is the introduction of regenerative payload capabilities, enabling 5G system functionalities directly onboard satellite platforms. Unlike the transparent payloads supported in earlier releases, regenerative payloads allow for more flexible and efficient NTN deployments. Furthermore, NR NTN evolves to support RedCap UEs.

Release 19 further optimizes 5G-Advanced to better accommodate XR applications. This includes enabling transmission and reception during gaps or restrictions caused by RRM measurements and RLC's acknowledged mode of operation. Additionally, Release 19 explores improvements to the PDCP and uplink scheduling mechanisms, particularly focusing on incorporating delay information. 3GPP also investigates techniques to support XR applications more efficiently, ensuring they meet the diverse and stringent QoS requirements associated with multi-modal XR use cases.

At the NG-RAN architecture level, 3GPP leverages AI/ML to address more use cases in Release 19. One new use case is AI/ML-based network slicing, where AI/ML is utilized to optimize resource allocation dynamically for different network slices. Another focused area is coverage and capacity optimization, with AI/ML being leveraged to adapt cell and beam coverage dynamically, a technique often referred to as cell shaping.

Other enhancements to existing features in Release 19 include:

- Sidelink: This work focuses on multi-hop UE-to-network sidelink relay for mission-critical communications, especially in public safety and out-of-coverage scenarios.
- Network energy savings: Key enhancements include on-demand SSB in SCell for connected-mode UEs configured with CA, on-demand system information

- block type 1 (SIB1) for UEs in idle and inactive modes, and adaptation of the transmission of common signals and channels.
- Multi-carrier enhancements: One enhancement is to use a single DCI to schedule multiple cells with different subcarrier spacing values or carrier types. Another enhancement is to use a single DCI to schedule multiple cells with multiple PDSCH/PUSCH transmissions on each cell.

### B. New Features

The Release-18 study on AI/ML for NR established a framework, as illustrated in Figure 6. Following this, Release 19 introduces support for one-sided AI/ML models, applicable to either the UE or network side [65]. Signaling and protocol aspects related to lifecycle management are specified, facilitating AI/ML-based beam management and positioning for both UE- and network-sided models, and AI/ML-based CSI prediction for UE-sided models. Besides these normative advancements, Release 19 continues to study CSI compression which relies on two-sided AI/ML models involving a UE-side encoder and a network-side decoder, as well as AI/ML for mobility.

SBFD introduces simultaneous downlink and uplink coexistence within the TDD carrier, blending TDD and FDD techniques [66]. By configuring an uplink subband during the time allocated for legacy downlink symbols in TDD, SBFD extends uplink duration, improving uplink coverage and latency [67]. Release 19 specifies the time and frequency locations of SBFD subbands and establishes procedures for transmission, reception, measurement, and random access in SBFD symbols. Additionally, Release 19 addresses CLI which arises when one node transmits while another receives on the same frequency band.

UE power usage can be reduced if the UE wakes up only when it detects an LP-WUS from the gNB. This approach involves equipping the UE with a low-power wake-up receiver (LP-WUR) alongside its main radio [68]. The LP-WUR activates the UE's main radio upon detecting an LP-WUS from the gNB. Release 19 specifies LP-WUS design and procedures [69]. For UEs in idle/inactive mode, configurations enable LP-WUS to activate the main radio for paging monitoring, relax serving and neighbor cell measurements for RRM, and offload serving cell RRM measurements to the LP-WUR. For UEs in connected mode, procedures are specified to allow LP-WUS to trigger the main radio for monitoring PDCCH.

Ambient IoT technology represents a significant leap in enabling ultra-low-power and battery-less IoT devices [70]. These devices operate by harvesting energy from ambient sources, such as radio waves, and use backscatter communication to transmit data. This approach reduces complexity and power consumption, making it ideal for applications like asset tracking and environmental monitoring. The Release-19 work item standardizes the ambient IoT framework, focusing on deployment in the licensed FR1 spectrum, with devices consuming as little as 1 µW of power [71]. It includes specifications for communication protocols, lightweight signaling, and RF requirements tailored to ambient IoT. This new feature expands the potential of cellular IoT, complementing existing 3GPP low-power wide-area technologies like NB-IoT [72] and LTE machine-type communication (LTE-M) [73].

Release 19 introduces wireless access backhaul (WAB) nodes to enhance 5G RAN topologies to support broader use cases, such as providing connectivity for onboard UEs in aircraft, ships, or vehicles in remote areas. A WAB node includes CU, distributed unit (DU), and mobile termination (MT) functions, unlike IAB nodes, which support only DU and MT functions. Key areas of focus include WAB architecture, protocol enhancements, and resource multiplexing.

5G femtocell extends the concept of LTE hone eNB, delivering NR access at home and enterprise premises [74]. It provides cost-effective indoor coverage and offloads macro gNB traffic. Release 19 specifies NR femtocell architecture with optional femto-gateway for NG interface and access control in open, hybrid, and closed modes.

### C. Toward 6G

Release 19 lays the groundwork for 6G by addressing critical areas that will serve as stepping stones toward 6G, including channel modeling for ISAC and the 7-24 GHz spectrum. These channel models will be critical for the design and evaluation of 6G systems.

ISAC is a key usage scenario of 6G networks, aiming to merge wireless communication with sensing capabilities [75]. ISAC allows the same infrastructure to support both data transfer and spatial sensing, which can detect and track objects. Release 19 initiates a comprehensive study on ISAC channel modeling, identifying deployment scenarios and refining channel models to support sensing. Key aspects of the ISAC channel modeling work include sensing target characterization, environment and mobility modeling, and spatial consistency.

The 7-24 GHz spectrum (aka., frequency range 3 (FR3)) has emerged as a key frequency range for 6G development, offering a balance between wide-area coverage and high data rates [76]. Release 19 focuses on validating and enhancing existing 3GPP channel models for this spectrum. Existing models are verified and refined using new measurement and ray tracing data to ensure their accuracy for this intermediate spectrum range. The larger antenna arrays expected in this frequency range require additional modeling considerations, such as near-field propagation effects and spatial non-stationarity [77]. These factors influence the accuracy of beamforming and MIMO operations, which are crucial for high-performance 6G networks.

## VII. 3GPP RELEASE 20: EVOLUTION TO 6G

3GPP Release 20, set to start in the second half of 2025, marks the third phase of 5G-Advanced evolution. Building on the foundational work of Releases 18 and 19, Release 20 will focus on addressing critical commercial needs, enhancing network performance, and preparing for the eventual transition to 6G. The release will manage a delicate balance between enhancing 5G-Advanced technologies and exploring forward-looking studies essential for developing 6G specifications in Release 21. While approvals of detailed study and work items are planned for June 2025, the 3GPP workshops in December 2024 provided direction for Release 20, underlining the dual goals of optimizing 5G-Advanced systems and preparing for the evolution to 6G [78].

As a cornerstone technology of 5G, MIMO is anticipated to continue to evolve in Release 20. The release will focus on enhancing downlink and uplink functionalities to address real-world commercial needs. After continuous enhancements since Release 15, Release 20 will selectively target high-impact MIMO topics, such as improved beamforming, enhanced channel reporting, and scalability for massive antenna arrays.

AI/ML is expected to continue to play a major role in 5G-Advanced evolution, with Release 20 building on the advancements of previous releases. A key focus will be on integrating AI/ML into mobility management, such as RRM prediction and event forecasting (e.g., handover failures and radio link failures). New AI/ML use cases for NG-RAN will likely be introduced. AI/ML for NR air interface also receives significant attention, particularly in the area of CSI feedback. Enhanced UE data collection capabilities will further advance the use of AI/ML.

Ambient IoT technology, which enables ultra-low-power and battery-less IoT devices, will remain a high-priority topic in Release 20. The work in this area will build on Release 19 advancements, aiming to support additional use cases, deployments, connectivity topologies, devices, and traffic types. For example, Release 20 will support a new device type characterized by a slightly higher complexity and peak power consumption (up to a few hundred μW) than the one considered in Release-19 ambient IoT.

Enabling voice calls over GEO satellites using NB-IoT technology has also attracted significant attention for Release 20. GEO satellites provide broad global coverage, making them ideal for emergency short message service and voice calls. Release 20 is expected to address the limitations of existing NB-IoT systems, which were initially designed for IoT use cases and do not support voice services [79]. This will expand the capabilities of GEO satellite communication and meet the growing demand for global, reliable voice services.

Sensing emerges as a key theme, with Release 20 potentially exploring its integration into 5G-Advanced systems. Targeting applications such as UAV navigation and smart transportation, the scope may include gNB-based mono-static sensing by utilizing existing NR physical layer design. The potential work will focus on minimizing the 5G specification impact and coordinate closely with related 6G initiatives to avoid duplication of standards development.

For the evolution to 6G, 3GPP has outlined a comprehensive timeline for 6G studies. The work begins with the use case and requirement studies, followed by technical studies in Release 20. The technical study phase will commence in Q3 2025, lasting approximately 21 months. This phase will explore various 6G technologies, evaluate their performance, and identify those most suitable for implementation. These studies will provide the foundation for the work item phase in Release 21, during which 3GPP will finalize the first 6G specifications. This process ensures that 3GPP remains on track to enable initial commercial 6G systems by 2030.

## VIII. WHAT SHOULD 6G STANDARDIZATION DO DIFFERENTLY?

In the previous sections, we have reviewed the evolution of 3GPP from Release 15 to Release 20. This decade-long evolution journey in 3GPP has provided valuable lessons for future mobile network standards and technologies. This section presents five key lessons, informed by standardization experience and intensive discussions with key players in the telecommunication industry. Table 2 summarizes the lessons and their implications for 6G.

*Lesson #1: Overly complicated 5G design with excessive options led to fragmentation and limited deployment of features.*

As reviewed in this retrospective, 5G evolution has unlocked a wealth of innovative ideas and features, aiming to address diverse use cases and deployment needs. However, the overly complicated 5G design, where multiple options are provided for each functionality, has resulted in fragmentation and inefficiencies. The flexibility increased complexity in device and network implementation and led to fragmented adoption, with some features only implemented for specific regions or industries, and high testing and deployment costs for vendors and operators. For example, sidelink evolution in 3GPP introduced multiple configurations and features across different 3GPP releases, but its commercial deployment is yet to take off.

**6G standardization should focus on technical excellence to achieve simplicity and scalability.** The 4G LTE standard achieved widespread success by emphasizing technical excellence and global alignment without excessive fragmentation. Standards for 6G should follow this example, aiming for high-quality solutions rather than compromising with overly flexible or fragmented approaches. Besides, standards should avoid adding excessive layers of complexity to features across releases. For enhancements, features that have clear use cases with high demand and significant commercial potential should be prioritized.

*Lesson #2: Focusing on impractical 5G performance requirements led to complex features that were not deployed.*

The focus on achieving extreme performance metrics in 5G, such as URLLC with sub-millisecond latency and near-perfect reliability, resulted in features that were technically impressive but not deployed. This is due to a number of factors, including the high complexity in implementation, lack of immediate commercial demand, and limited practical use cases that justify the cost and effort of deployment. Furthermore, pursuing air interface latency reductions below 1 ms as a performance metric for 5G offered minimal practical benefits because other aspects, such as transport latency and application-layer latency, become dominant factors shaping the user experience at this level. Similarly, the heavy emphasis on peak data rates (at least 20 Gbps in downlink and 10 Gbps in uplink) set unrealistic expectations about 5G capabilities. The theoretical peak data rates rarely translated to real-world experiences for most users.

**6G standardization should focus on realistic and meaningful performance improvements.** Instead of over-optimizing for extreme scenarios, 6G should focus on requirements that are grounded in real-world demand. It is more meaningful to prioritize broader user experience metrics, such as joint rate-latency-reliability requirements. Additionally, instead of emphasizing peak data rates, 6G should deliver meaningfully improved user-experienced data rates, particularly at lower-percentile performance levels.

| Lessons Learned from 5G Evolution | What Should 6G Standardization Do Differently |
|---|---|
| Lesson #1: Overly complicated 5G design with excessive options led to fragmentation and limited deployment of features. | 6G standardization should focus on technical excellence to achieve simplicity and scalability. |
| Lesson #2: Focusing on impractical 5G performance requirements led to complex features that were not deployed. | 6G standardization should focus on realistic and meaningful performance improvements. |
| Lesson #3: Multiple architecture choices in 5G created unnecessary complexity and delayed adoption. | 6G standardization should focus on standalone architecture from the start. |
| Lesson #4: Bit transport alone limits telcos to "dumb pipes" and restricts growth opportunities. | 6G standardization should focus on integrating computing, AI, and sensing to enable multi-tenancy. |
| Lesson #5: The traditional 3GPP release-style enhancements have difficulty keeping up with the increasing pace of innovation. | 6G standardization should enable continuous enhancements and innovations. |

Table 2: A summary of lessons learned from 5G evolution and their implications for 6G.

*Lesson #3: Multiple architecture choices in 5G created unnecessary complexity and delayed adoption.*

3GPP specified many architecture options for combining 4G LTE and 5G NR, with only two options, NSA and SA, eventually deployed in commercial networks. NSA mode was designed to accelerate early 5G rollouts, leveraging the 4G core network. However, it required tight interworking between 4G and 5G components. SA mode enabled 5G to operate with its own 5G core network, unlocking the full capabilities of 5G, but deploying SA required a more significant infrastructure overhaul and took longer to roll out. As a result, operators had to deploy NSA first to meet market demands for 5G services quickly, then transition to SA later. This dual-phase rollout resulted in increased costs for operators, fragmentation in the ecosystem, and delayed realization of SA-specific benefits.

**6G standardization should focus on standalone architecture from the start.** A single standalone mode ensures that 6G is deployed with its dedicated core network and RAN from the outset, avoiding transitional phases and complexities. Operators can focus resources on deploying one robust architecture, reducing the costs and operational overhead associated with dual-mode support. Device manufacturers and vendors can concentrate efforts on a single ecosystem, leading to faster development cycles, simplified testing, and accelerated adoption.

*Lesson #4: Bit transport alone limits telcos to "dumb pipes" and restricts growth opportunities.*

5G enhancements have largely concentrated on improving throughput, latency, and spectral efficiency. Relying solely on bit transport confines telecommunications operators to the role of "dumb pipes," where their networks merely facilitate data flow without capturing value from the services or insights that ride on them. It sidelines operators from higher-value ecosystems dominated by cloud providers, over-the-top players, and application developers. As a result, this model leads to the commoditization of operators' business and restricts revenue growth. This, in turn, has made some operators hesitant to invest in 5G infrastructure.

**6G standardization should focus on integrating computing, AI, and sensing to enable multi-tenancy.** In the 6G era, mobile networks will merge with cloud services to provide compute-as-a-service, enabling a connected, intelligent edge with AI and communications running on a shared infrastructure, i.e., multi-tenancy. Future networks must evolve to become platforms for data production, processing, and consumption, enabling operators to extract value beyond pure data transport. The integration of sensing capabilities allows networks to act as sensors, detecting and interpreting data such as object movement, environmental changes, and user contexts. The combination of computing, AI, and sensing will enable the creation of digital twins, providing a unique and valuable asset for operators [80].

*Lesson #5: The traditional 3GPP release-style enhancements have difficulty keeping up with the increasing pace of innovation.*

As seen in the previous sections of this retrospective, 3GPP evolution depends on the release cycle, where enhancements are bundled and introduced every 18-24 months. Fixed schedules for 3GPP releases have the drawback of delaying the deployment of new features, forcing the industry to wait for standardization processes. This release-cycle evolution approach is increasingly misaligned with the rapid pace of AI-driven innovation. AI technologies are advancing rapidly, often delivering significant breakthroughs within months or even weeks. The rigid cadence of 3GPP releases means that AI-driven capabilities may be delayed in reaching deployment. Waiting for the next release cycle undermines AI's ability to optimize network performance.

**6G standardization should enable continuous enhancements and innovations.** 6G should start to gradually move away from release-bound updates, enabling innovations to be deployed dynamically as they are developed. Shifting to an AI-native system design in 6G will enable the system to evolve dynamically through self-learning and adaptation, facilitating continuous improvements without waiting for standardization cycles. Additionally, software-driven enhancements can reduce deployment costs and time-to-market for innovations.

## IX. CONCLUSION AND FUTURE OUTLOOK

The evolution of 3GPP from Release 15 to Release 20 has been a decade-long journey of innovation and collaboration, transforming the promise of 5G into reality and setting the stage for 6G. Each release has contributed uniquely to building a robust and versatile ecosystem that has advanced connectivity, expanded possibilities, and prepared the groundwork for the next generation of mobile networks. One of the most significant takeaways from this decade-long journey is the importance of adaptability and collaboration. 3GPP's ability to align diverse stakeholders has been instrumental in driving consensus and progress.

As we look ahead to the 6G era, the lessons learned and foundations built over the past ten years will serve as invaluable guideposts [81]. 6G envisions a hyper-connected world where communication, sensing, and AI converge to create seamless and immersive experiences. The journey to 6G is already underway. The formal 6G requirements process, led by ITU in collaboration with 3GPP, will define the benchmarks for IMT-2030 technologies. These requirements will encompass not only traditional metrics such as data rate, latency, and reliability but also new dimensions like sustainability, sensing, and AI. The convergence of AI, edge computing, and cloud technologies will further enhance the adaptability and intelligence of 6G networks, enabling them to anticipate and respond to user needs in real-time. With continued dedication to innovation, inclusivity, and sustainability, 6G has the potential to reshape our world in ways we are only beginning to imagine.